\documentclass[12pt]{article}
\usepackage{graphicx}
\usepackage{amsmath}
\usepackage{amssymb}
\usepackage{authblk}
\usepackage[left=0.9in,right=0.9in,top=0.8in,bottom=0.8in]{geometry}
\usepackage{setspace}
\usepackage{float}
\usepackage[super,sort&compress,comma]{natbib}
\usepackage[version=4]{mhchem} 
\usepackage{graphicx} 
\usepackage[font={normalsize}]{caption} 
\usepackage[labelfont=bf]{caption}
\usepackage{amsmath, amssymb}
\usepackage{natbib}
\usepackage{booktabs}
\usepackage{array}
\usepackage{subcaption}
\usepackage{multirow}
\usepackage{soul}
\usepackage{float}
\usepackage{siunitx}
\usepackage{tabularx}
\usepackage{ulem}
\usepackage{hyperref} 
\usepackage{xcolor}
\usepackage{url}
\usepackage[labelfont=bf]{caption}
\usepackage[font={small}]{caption} 

\usepackage{tikz}
\usetikzlibrary{positioning, fit, calc, shapes.geometric, arrows.meta, backgrounds, shadows, intersections, spath3}

\begin{document}

\title{\textbf{Machine-Learned Leftmost Hessian Eigenvectors for Robust Transition State Finding}}

\author{Guanchen Wu$^{1}$, Chung-Yueh Yuan$^{1,5}$, Kareem Hegazy$^{4,7}$, Samuel M. Blau$^{6}$, and Teresa Head-Gordon$^{1-3,5}$}
\date{}
\maketitle

\begin{center}
\vspace{-24pt}
$^1$Kenneth S. Pitzer Theory Center and Department of Chemistry, $^2$Department of Bioengineering, $^3$Department of Chemical and Biomolecular Engineering, $^4$Department of Statistics, University of California, Berkeley, CA, 94720 USA\\
$^5$Chemical Sciences Division and $^6$Energy Technologies Area, Lawrence Berkeley National Laboratory, Berkeley, CA, 94720 USA\\
$^7$International Computer Science Institute, Berkeley, CA 94704 USA\\

corresponding author: thg@berkeley.edu
\end{center}

\begin{abstract}
\noindent
The reliable determination of transition states (TSs) benefits from second-order information  for robust convergence and validation, but the computational expense of Hessians prohibits their routine use in TS optimization. Here, we present a machine-learning-driven TS optimizer that directly predicts the leftmost Hessian eigenvector (LMHE), the critical mode that locally approximates the reaction coordinate encompassing the TS. We demonstrate that our LMHE optimizer recovers TS solutions at the same rate as full Hessian optimizers, and more robustly from degraded initial guess geometries, thereby eliminating the excessively long wall times characteristic of full-Hessian approaches and reducing total gradient evaluations compared to standard quasi-Newton methods. We further improve accuracy and robustness using uncertainty quantification for identifying occasional LMHE prediction failures, that then falls back to a full Hessian update from the machine learned potential at that optimization step, avoiding expensive active learning. Overall our methodology and semi-automated workflow delivers second-order stability at first-order computational expense to provide a highly efficient engine for high-throughput reaction discovery. 

\end{abstract}

\section{Introduction}
Understanding reaction chemistry and predicting kinetics requires a detailed knowledge of the potential energy surface (PES), especially the location and characterization of transition states (TSs). Mathematically defined as first-order saddle points on the PES, TSs represent the highest energy point along the minimum energy path connecting reactants to products of an elementary reaction. They are characterized by having exactly one negative eigenvalue in the Hessian matrix, corresponding to a single imaginary vibrational frequency. Unlike equilibrium structures which can be determined by optimizing along descent directions, locating saddle points is a more complex optimization as it requires maximizing the energy along a single specific mode, the reaction coordinate, while simultaneously minimizing it along all other degrees of freedom.\par

The central challenge in TS optimization is correctly identifying this single ascent direction. The success of a TS optimization often depends crucially on the quality of the curvature information from the Hessian matrix, as it provides the necessary geometric guidance to distinguish the reaction coordinate from orthogonal conformational or non-reactive modes. Second-order methods utilize the full Hessian matrix to explicitly identify the mode with the most negative curvature as the local approximation of the reaction coordinate. While robust, calculating the exact Hessian at every optimization step using \textit{ab initio} methods such as density functional theory (DFT) is prohibitively expensive for most practical applications. 

Consequently, modern geometry optimization on the \textit{ab initio} PES typically relies on quasi-Newton (QN) methods to construct cheaper approximate Hessians using only gradient information. While the Broyden-Fletcher-Goldfarb-Shanno (BFGS) algorithm is arguably the most widely used approach for minimization, its standard formulation enforces a positive-definite Hessian, making it inherently unsuitable for locating saddle points which require indefinite Hessians \cite{bfgs1, bfgs2, bfgs3, bfgs4}. To address this, other update schemes capable of handling indefinite curvature have been developed, such as Powell-symmetric Broyden (PSB) \cite{psb1, psb2}, symmetric rank one (SR1) \cite{sr1}, Murtagh-Sargent-Powell (MSP) \cite{msp1, msp2} and TS-BFGS \cite{tsbfgs1, tsbfgs2} methods \cite{qnmethods}. These QN methods attempt to recover curvature information solely from the optimization trajectory history (i.e. coordinate and gradient changes). However, this reconstruction is often inaccurate, particularly at the beginning of the optimization or in flat regions, leading to convergence to an unintended TS rather than the target reaction-product pair, or failure to find the saddle point at all.\par 

To mitigate this, more advanced QN strategies have been developed to better recover the transition mode. Some approaches calculate an exact full Hessian at the first step to ensure a good starting point, though this information decays over time. Others periodically employ iterative diagonalization (e.g. Jacobi-Davidson) to actively locate the leftmost eigenvector during the run \cite{jd1, jd2, jd3, jd4, Sella2}. While this strategy represents a significant improvement over history-only methods, robustness remains a critical challenge. Furthermore, iterative diagonalization introduces a new computational cost: determining the eigenvector via finite difference requires multiple auxiliary gradient evaluations. This cost prevents applying diagonalization at every optimization step such that standard optimizers apply diagonalization only intermittently and rely on history-based updates in between, limiting robust and rapid convergence. The reliance on these fragile TS optimizations also limits the efficacy of broader reaction path search workflows as well. Even when employing established double- and single-ended interpolation methods such as the Nudged Elastic Band (NEB)\cite{neb1, neb2}, Quadratic Synchronous Transit (QST)\cite{qst}, or the Growing String Method (GSM)\cite{gsm}, a subsequent local optimization is typically required to pinpoint the exact saddle point. As a result, despite the availability of these global search strategies, the final location of the TS often remains a bottleneck, requiring significant user involvement and trial-and-error when Hessian information is absent.\par

Recent advances in geometric deep learning have presented a potential solution to TS optimization. Machine learning interatomic potentials (MLIPs) have demonstrated the ability to achieve near quantum-chemical accuracy, and as demonstrated by Yuan et al., the analytical full Hessian can be obtained from these potentials via automatic differentiation, offering a significant computational speedup over \textit{ab initio} methods \cite{Yuan2024}. Nevertheless, this second-order MLIP approach is not without computational bottlenecks. The calculation of exact second-order derivatives via automatic differentiation necessitates the construction and retention of extensive computational graphs. While feasible, this process remains significantly more computationally intensive than simple gradient evaluations, imposing a substantial memory and time overhead if performed at every optimization step.\par

In this work, we propose a novel ML architecture and TS optimization strategy to solve this key dilemma between robustness and computational efficiency. Our approach leverages the fundamental insight that the critical curvature information required for efficient TS optimization is encapsulated primarily in the leftmost Hessian eigenvector (LMHE), the vector corresponding to the most negative eigenvalue, which locally approximates the reaction coordinate and defines the ascent direction for TS optimization. We train a MLIP model to predict this vector directly from coordinates. But because TS modes often involve concerted, non-local atomic motions, they pose a unique challenge for learning the LMHE using standard local message-passing neural networks (MPNNs). To capture these motional dependencies efficiently, we introduce a novel $E(3)$-equivariant architecture that couples a message passing encoder with an induced global attention decoder that avoids the prohibitive quadratic scaling of standard full attention. The resulting learned LMHE is used to guide the geometry optimization within a restricted-step partitioned rational function optimization (RS-PRFO) framework \cite{prfo1, prfo2, prfo3}, and failure modes identified through uncertainty quantification provides an occasional correction using a full Hessian update to improve success rates. Our LMHE approach is implemented in the user-friendly Sella optimization software package\cite{Sella}, and is tested across 240 organic combustion from the Sella benchmark set \cite{Sella}. We show our method achieves the robustness of second-order methods at a fraction of the computational cost, thereby replacing expensive full Hessian calculations or iterative diagonalization at each optimization step, aided by a workflow that semi-automates the success in TS convergence.\par

\section{Methods}

\subsection{Data Preparation for energies, gradients, Hessians, and LMHE} \label{data_preparation}
All energy and gradient evaluations were conducted on the PES defined by the NewtonNet MPNN model\cite{NewtonNet} trained on conformations from the original Transition1x (T1x) dataset\cite{T1x}, augmented with one million structures from the ANI-1x dataset \cite{ANI1x}, and described in previous work\cite{Yuan2024}. This choice of MLIP is motivated by two factors. First, the NewtonNet model is fine-tuned on an augmented T1x dataset which contains extensive reactivity data, explicitly sampling reaction pathways and saddle points rather than just equilibrium geometries. As demonstrated by Yuan et al., this specialized training regime accurately reproduces the topography of the reference DFT PES around TSs, making the NewtonNet MLIP model highly effective for TS finding \cite{Yuan2024}. Second, the differentiable nature of the NewtonNet MLIP enables the calculation of exact Hessians via automatic differentiation, and allows us to include a rigorous full Hessian baseline in our benchmarks, which would otherwise be prohibitively expensive to compute with \textit{ab initio} methods such as DFT.\par

The dataset for training the LMHE predictors was generated using the fine-tuned NewtonNet model\cite{NewtonNet, Yuan2024}. For each structure in the training data, the full Hessian was computed via automatic differentiation\cite{Yuan2024}. The six (or five for linear molecules) trivial eigenvectors corresponding to overall translation and rotation were removed through projection. The LMHE was then identified from the remaining non-trivial modes. To prevent data leakage, the dataset was partitioned into training, validation, and test sets based on chemical composition, strictly following the partitioning strategy used for the NewtonNet fine-tuning\cite{Yuan2024} for consistency. \par

\subsection{Eigenvector-Informed Hessian Update Scheme} \label{hessian_update}

To integrate the machine-learned LMHE into the TS optimization, a modified QN Hessian updating scheme was developed. This approach leverages the predicted eigenvector to construct a more accurate approximate Hessian at each optimization step. The geometry updates were then performed within a RS-PRFO framework \cite{prfo1, prfo2, prfo3}, as implemented in the Sella optimizer \cite{Sella2}.\par

In the notation that follows, all vectors are assumed to be column vectors. At optimization step $k$, the approximate Hessian for the subsequent step, $\mathbf{H}^{(k+1)}$, is constructed by incorporating the predicted leftmost eigenvector, $\mathbf{v}_1^{(k+1)}$. The Hessian is decomposed into components parallel ($\parallel$) and perpendicular ($\perp$) to this eigenvector:
\begin{equation}
    \mathbf{H}^{(k+1)}=\mathbf{H}_{\parallel}^{(k+1)}+\mathbf{H}_{\perp}^{(k+1)}.
\end{equation}
The parallel component is constructed from the eigenvector and its corresponding eigenvalue, $\lambda_1^{(k+1)}$:
\begin{equation}
    \mathbf{H}_{\parallel}^{(k+1)}=\lambda_1^{(k+1)}\frac{\mathbf{v}_1^{(k+1)} {\mathbf{v}_1^{(k+1)}}^{\mathrm{T}} }{{\mathbf{v}_1^{(k+1)}}^{\mathrm{T}} {\mathbf{v}_1^{(k+1)}}}.
\end{equation}
The leftmost eigenvalue, $\lambda_1^{(k+1)}$, is estimated using the Rayleigh-Ritz quotient:
\begin{equation}
    \lambda_1^{(k+1)}=\frac{\mathbf{y}^\mathrm{T}{\mathbf{v}_1^{(k+1)}}}{\mathbf{s}^\mathrm{T}{\mathbf{v}_1^{(k+1)}}},
\end{equation}
where $\mathbf{s}=\mathbf{x}^{(k+1)}-\mathbf{x}^{(k)}$ is a vector representing the change in atomic coordinates and $\mathbf{y}=\mathbf{g}^{(k+1)}-\mathbf{g}^{(k)}$ is the corresponding vector representing the change in gradients. \par

The perpendicular component of the Hessian is updated in the subspace orthogonal to $\mathbf{v}^{(k+1)}_{1}$. This is achieved using a projection matrix, $\mathbf{P}_{\perp}$:
\begin{equation}
    \mathbf{P}_{\perp}=\mathbf{I}-\frac{{\mathbf{v}_1^{(k+1)}} {\mathbf{v}_1^{(k+1)}}^\mathrm{T}}{{\mathbf{v}_1^{(k+1)}}^\mathrm{T} {\mathbf{v}_1^{(k+1)}}},
\end{equation}
where $\mathbf{I}$ is the identity matrix. The perpendicular components of the coordinate change ($\mathbf{s}_{\perp}$), gradient change ($\mathbf{y}_{\perp}$), and the previous Hessian ($\mathbf{H}_{\perp}^{(k)}$) are then calculated:
\begin{equation}
    \mathbf{s}_{\perp}=\mathbf{P}_{\perp}\mathbf{s},\quad
    \mathbf{y}_{\perp}=\mathbf{P}_{\perp}\mathbf{y},\quad
    \mathbf{H}^{(k)}_{\perp}=\mathbf{P}_{\perp}\mathbf{H}^{(k)}\mathbf{P}_{\perp}.
\end{equation}\par

By explicitly partitioning out the transition mode (the $\mathbf{H}_{\parallel}$ component), the optimization problem in the remaining ($N-1$) dimensional subspace is effectively a minimization problem, for which QN updates are well-established and robust. Thus, the perpendicular Hessian component is updated using a standard QN formula:

\begin{equation} \label{perpendicluar_update}
    \mathbf{H}_{\perp}^{(k+1)} = \mathrm{QN}(\mathbf{H}_{\perp}^{(k)}, \mathbf{s}_{\perp}, \mathbf{y}_{\perp}).
\end{equation}

While Eq. \ref{perpendicluar_update} is compatible with various QN formulations, this work employs the TS-adapted Broyden-Fletcher-Goldfarb-Shanno method (TS-BFGS) \cite{tsbfgs1, tsbfgs2}. Unlike the standard BFGS update, which requires a positive-definite Hessian, the TS-BFGS formulation is designed to be compatible with indefinite approximate Hessians. This provides the necessary flexibility for the perpendicular components to possess negative eigenvalues, thereby allowing the full approximate Hessian to temporarily contain multiple negative eigenvalues during optimization. A complete mathematical description of this update is provided in Supplementary Appendix 1.

To ensure that the predicted eigenvector corresponds to the lowest eigenvalue of the composite Hessian $\mathbf{H}^{(k+1)}$, the estimated eigenvalue $\lambda_1^{(k+1)}$ is constrained. If $\lambda_1^{(k+1)}$ is greater than the lowest eigenvalue of the perpendicular Hessian ($\lambda_{\perp1}^{(k+1)}$), it is adjusted as follows:
\begin{equation}
    \lambda_1^{(k+1)}=\lambda_{\perp1}^{(k+1)}-1.
\end{equation}\par

For the first optimization step, we initialize the Hessian as:
\begin{equation}
    \mathbf{H}^{(0)} = \mathbf{I} - 2 \frac{\mathbf{v}_1^{(0)} {\mathbf{v}_1^{(0)}}^\mathrm{T} }{{\mathbf{v}_1^{(0)}}^\mathrm{T} {\mathbf{v}_1^{(0)}}}.
\end{equation}
This specific construction for the initial Hessian is chosen to precondition the optimizer by immediately incorporating the machine-learned leftmost eigenvector. The resulting $\mathbf{H}^{(0)}$ matrix has an eigenvalue of $-1$ along the predicted eigenvector $\mathbf{v}_1^{(0)}$ and eigenvalues of $+1$ for all modes in the orthogonal subspace. This initialization provides the RS-PRFO framework with a Hessian that already possesses the correct qualitative structure of a TS (i.e. one negative mode) and aligns this mode with the predicted leftmost eigenvector, offering a significant advantage over a default positive-definite guess (e.g. the identity matrix) which lacks any imaginary frequency.

\subsection{Transition State Optimization} \label{ts_optimization}
The performance of the TS optimizers is evaluated by the Sella benchmark dataset\cite{Sella}. This dataset contains 500 organic molecules (7 to 25 atoms) in configurations that approximate TS geometries of different reactions, among which 265 are closed-shell. As 25 of these reactions are also present in the T1x dataset, they were excluded from our analysis leaving 240 reactions for analysis. 

All TS optimizations and intrinsic reaction coordinate (IRC) calculations for the Sella benchmark dataset were performed in Cartesian coordinates using the Sella optimizer\cite{Sella2}, implemented using the Atomic Simulation Environment (ASE) \cite{ASE}. These optimizations were conducted on the NewtonNet PES, defined by the same model used for the augmented T1x data preparation (Section \ref{data_preparation}). The Sella implementation was modified to allow for the provision of an external Hessian matrix at each optimization step.\par

The Sella framework involves iterative diagonalization using the Rayleigh–Ritz procedure with a modified Jacobi–Davidson method (JD0, or Olsen’s method), a finite difference step size of $10^{-4}$ \r{A}, and a convergence threshold of 0.1 \cite{Sella2}. The trust radius of RS-PRFO is initialized to 0.1 and adjusted based on the ratio between the predicted and actual energy changes; the radius is increased by a factor of 1.15 when this ratio is below 1.035 and decreased by a factor of 0.65 when it exceeds 5.0. The IRC is located by performing energy minimization within a trust radius of $0.1\text{ \r{A}/amu}^{-1/2}$ in mass-weighted coordinates\cite{IRC}. A maximum of 1000 steps was imposed for both the TS optimization and the IRC search. To evaluate the optimization results, reactants and products were compared by testing for graph isomorphism. Molecular connectivity graphs, including atom indexing, were generated using Open Babel\cite{OpenBabel} and compared using the VF2 algorithm \cite{vf2}.

\subsection{Ensemble Consistency Check} \label{ensemble_consistency}
We employ an ensemble-based consistency check to govern a  fallback mechanism to the exact Hessian from automatic differentiation when prediction uncertainty is high. To facilitate this ensemble approach, we trained five independent instances of the 10.1 M parameter model with our proposed architecture, initialized with distinct random seeds, while the single-model optimizer utilizes one of these instances. The specific training details and hyper-parameters are provided in Supplementary Table S1. 

At each optimization step, the LMHE is predicted by all $N=5$ models in the ensemble, denoted as $\{\mathbf{v}_1^{(1)}, \dots , \mathbf{v}_1^{(N)}\}$. Each predicted vector is normalized to unit length. The final predicted eigenvector, $\bar{\mathbf{v}}_1$, is computed as the normalized mean of these ensemble members. Due to the sign ambiguity of eigenvectors (i.e. $\mathbf{v}$ and $-\mathbf{v}$ are equivalent), we quantify consensus using a sign-invariant metric based on the mean outer product matrix $\bar{\mathbf{Q}}$:
\begin{equation}
    \bar{\mathbf{Q}}=\frac{1}{N}\sum_{i=1}^{N} \mathbf{v}_1^{(i)} \mathbf{v}_1^{(i)^{\mathrm{T}}}.
\end{equation}
The prediction uncertainty, $\sigma$, is defined as:
\begin{equation}
    \sigma=1-\lambda_{max}(\bar{\mathbf{Q}}),
\end{equation}
where $\lambda_{max}(\bar{\mathbf{Q}}$ is the largest eigenvalue of $\bar{\mathbf{Q}}$. This formulation naturally accounts for the sign ambiguity, as the outer product $\mathbf{v}\mathbf{v}^{\mathrm{T}}$ is invariant to the transformation $\mathbf{v}\mapsto -\mathbf{v}$. The uncertainty metric $\sigma$ ranges from 0 (perfect consensus, where all predictions are collinear) to a value approaching 1 (maximal disagreement).\par

If $\sigma$ exceeds a predefined threshold $\tau$, the prediction is considered unreliable. In such cases, the optimizer reverts to calculating the exact Hessian via automatic differentiation for that specific step. Conversely, if $\sigma$ is below $\tau$, the normalized mean vector $\bar{\mathbf{v}}_1$ is utilized as the consensus prediction in the update scheme described in Section \ref{hessian_update}.

\section{Results}

\subsection{GotenNet and Global Attention Architecture}
\begin{figure}[t!]
\centering
\includegraphics[width=0.99\textwidth]{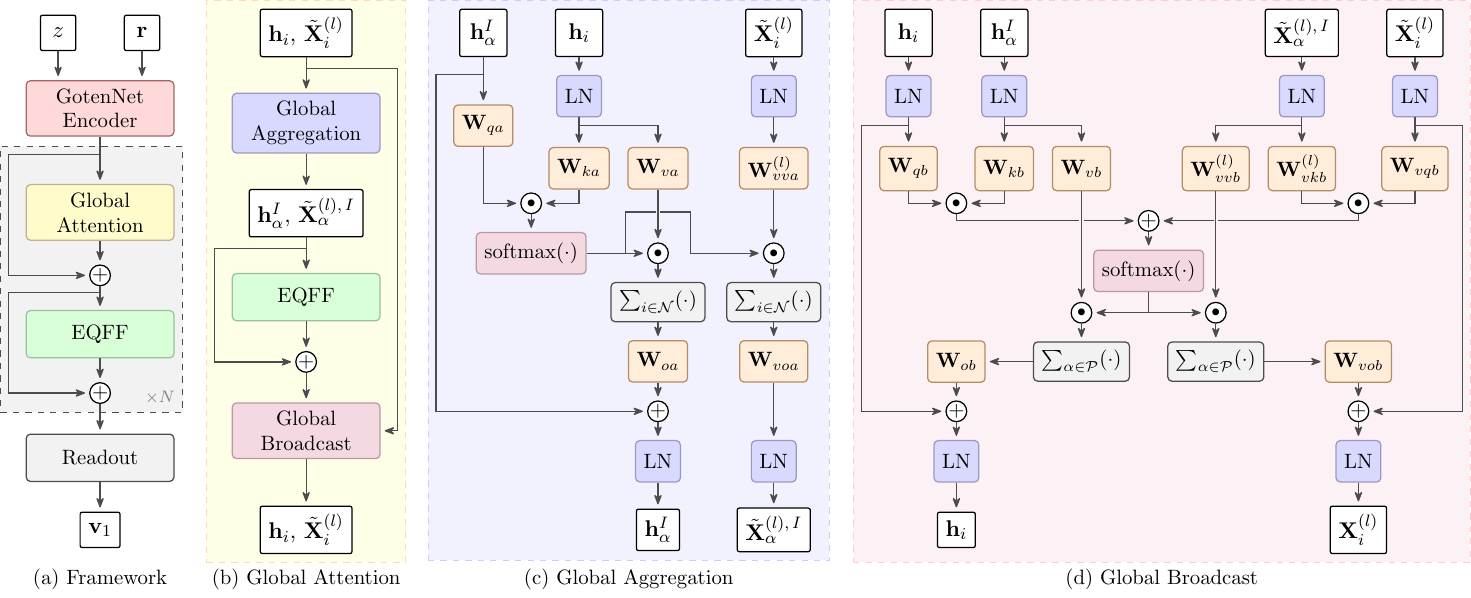}
    \caption{\textit{GotenNet-GA model architecture used for transition state searches with the learned LMHE.} (a) The macroscopic framework, comprising a GotenNet encoder, a global interaction module, and a readout. (b) Architecture of the Global Attention block, illustrating its internal mechanism and explicitly incorporating the encompassing residual connection shown in (a). (c) The Global Aggregation block which maps features from the physical graph nodes to the latent inducing points. (d) The Global Broadcast block, which propagates the updated global context from inducing points back to the physical nodes. In the figure, $+$ denotes addition, $\cdot$ denotes dot product, $\sum$ denotes aggregation with summation, $\mathcal{N}$ denotes the set of all real nodes in the graph, $\mathcal{P}$ denotes the set of all inducing points, and LN denotes layer normalization.}
    \label{fig:model}
\end{figure}

The prediction of the LMHE is geometrically analogous to predicting atomic forces, as both tasks require the output to transform equivariantly with the rotation and translation of the molecule. Our method utilizes an $E(3)$-equivariant neural network trained to predict the LMHE directly from atomic coordinates using a sequential encoder-decoder architecture shown in Figure \ref{fig:model}. 

The encoder, GotenNet \cite{GotenNet}, serves as an efficient equivariant MPNN to capture spatial information without the computational overhead often associated with irreducible representations connected via Clebsch-Gordon (CG) coefficients to ensure that molecular features rotate correctly with the molecule. While theoretically rigorous, these tensor product operations are computationally intensive and scale poorly, while GotenNet employs an attention mechanism that utilizes inner products of high-degree steerable features to capture angular dependencies and spatial relationships instead of explicit CG transforms. This approach allows the network to learn rich, equivariant representations essential for predicting vector quantities like the LMHE while maintaining a small computational overhead. This efficiency is particularly critical for our iterative TS optimization workflow, where the model must be inferred repeatedly to guide the geometry updates.\par

While atomic forces are largely determined by local environments, the LMHE often describes a collective motion of atoms across the entire molecule and is therefore an inherently non-local feature. Mathematically, this non-locality is a consequence of eigen-decomposition. Standard MPNN-based MLIPs, which rely on the assumption of locality to update atomic features, are constrained in capturing the long-range dependencies necessary to describe this concerted motion. To bridge this gap, we developed an architecture that couples the GotenNet encoder with a specialized $E(3)$-equivariant global attention decoder (Figure \ref{fig:model}). Unlike standard MPNNs that diffuse information slowly through local neighbors, this decoder leverages an induced set attention mechanism to aggregate local atomic features into a set of global inducing points, allowing the model to effectively capture the context in the entire molecule before broadcasting it back to the atomic level \cite{SetTransfromer}. This architecture ensures that the prediction is informed by global context without incurring the quadratic computational cost of conventional full attention mechanisms while ensuring that the entire architecture is $E(3)$-equivariant. The complete mathematical formulation of this decoder architecture is provided in Supplementary Appendix 2, and a formal proof demonstrating its $E(3)$-equivariance is presented in Supplementary Appendix 3. 

\subsection{Machine-Learned Leftmost Hessian Eigenvectors}
We trained two variants of our GotenNet-GA model with approximately 2.1 M and 10.1 M parameters on an augmented T1x\cite{T1x} dataset described in Methods Section \ref{data_preparation}. To evaluate the efficacy of the proposed architecture, we also trained two baseline models using the standard GotenNet architecture, explicitly constructed to match these parameter counts. The specific training details and hyper-parameters are provided in Supplementary Table S1. This controlled comparison allows us to isolate performance gains attributable to the global attention decoder, distinct from general scaling effects. We evaluated the errors of the models using the root mean square (RMS) sine value between the predicted and ground truth eigenvectors. Sine values are chosen to account for the sign ambiguity of eigenvectors, where 0 indicates perfect alignment and 1 indicates orthogonality (the worst possible alignment).


\begin{table}[H] 
    \centering
    \caption{\textit{Root mean square sine value for leftmost Hessian eigenvectors on the augmented T1x dataset.} The comparison includes baseline GotenNet models and our proposed architecture with global attention (GotenNet-GA) at both Small (S, $\sim$2.1 M parameters) and Medium (M, $\sim$10.1 M parameters) parameter scales.} \label{tab:loss}
    \begin{tabular}{lcccc}
        \toprule
         Subsets  & GotenNet$_{\text{S}}$ & GotenNet$_{\text{M}}$ & GotenNet-GA$_{\text{S}}$ & GotenNet-GA$_{\text{M}}$ \\
        \midrule
        Train  & \underline{0.33} & \textbf{0.26} & 0.37 & 0.29 \\
        Validation & 0.52 & \underline{0.48} & 0.50 & \textbf{0.46}  \\
        Test & 0.53 & \underline{0.49} & 0.51 & \textbf{0.47}  \\
         \bottomrule
     \end{tabular}
 \end{table}

The quantitative benefit of explicitly modeling global context is evident in Table \ref{tab:loss}. At the small scale ($\sim$2.1 M parameters), the  GotenNet-GA$_{\text{S}}$ architecture reduces the test error, and this performance advantage persists as model capacity increases at the medium scale, GotenNet-GA$_{\text{M}}$ with ~10.1 M parameters, achieving the lowest RMS error of 0.47. Notably, the gap between training and test performance is consistently narrower for our proposed GotenNet-GA architecture, supporting the idea that explicit modeling of global context facilitates more robust generalization to unseen reactions.\par

\subsection{Transition State Optimizations using LMHE}
We integrated the learned LMHE into the Sella optimizer \cite{Sella2} using the RS-PRFO framework \cite{prfo1, prfo2, prfo3}. Within the PRFO framework the leftmost eigenvector of the Hessian $\mathbf{v}_1$ locally approximating the reaction coordinate serves as the ascending direction along the PES for the TS search. This predicted vector $\mathbf{v}_1$ is utilized into a specialized Hessian update scheme (detailed in Section \ref{hessian_update}), providing a rigorous curvature guidance at every step of optimization.\par

To systematically evaluate the LMHE approach, we benchmarked the TS optimizer on 240 reactions from the Sella benchmark set \cite{Sella}. The initial TS geometries for these reactions were regenerated using KinBot with reaction templates \cite{KinBot}, which also define the intended reactant and product states. After each optimization, we followed the intrinsic reaction coordinate (IRC) from the resulting TS to determine the minimum energy path connecting the reactant and product wells. Outcomes are classified as intended (the IRC connects the exact KinBot target states), partially intended (only one of the two target state is matched), unintended (a valid TS is found, but connects to an alternative pathway), no reaction (optimization collapsed to a local minimum), or TS error (failure to converge). 

\begin{figure}[H]
\centering
\includegraphics[width=0.9\textwidth]{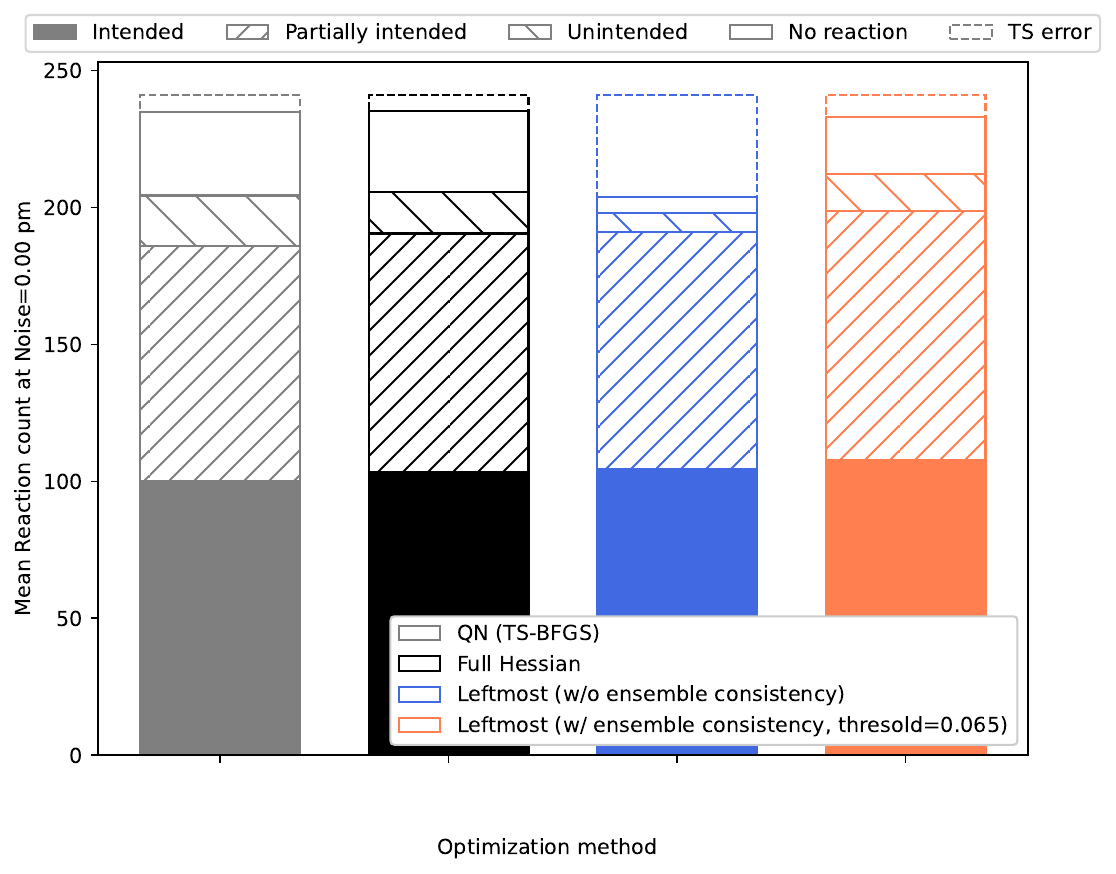}
  \caption{\textit{Comparison of predictions of different TS optimizers against 240 Sella benchmark reactions.} The performance of the LMHE approach against the general QN baseline (TS-BFGS) and full Hessian TS optimization methods is quantified by comparing the predicted reaction paths obtained through IRC calculations against the exact reactant and product states defined by KinBot templates. Outcomes are classified as intended (the IRC connects the exact target states), partially intended (only one target state is matched), unintended (a valid TS is found, but connects to an alternative pathway), no reaction (optimization collapsed to a local minimum), or TS error (failure to converge). The single model LMHE strategy exhibits a higher failure rate due to inaccurate LMHE predictions. The LMHE ensemble consistency check identifies unreliable predictions, triggering a fallback to exact Hessian calculations and reducing failure rates to levels that are competitive with the performance of the full Hessian optimization, and significantly outperforming the standard QN baseline.}
    \label{fig:intend_count}
\end{figure}

We compare our single inference LMHE method against two established baselines: the standard algorithm implemented in Sella\cite{Sella2}, which serves as a  state-of-the-art QN method  using iterative diagonalization, and a Full Hessian reference method where the exact Hessian is computed at every step via automatic differentiation of the MLIP. Our single inference LMHE approach demonstrates performance for intended and partially intended reactions as good as the Hessian and the QN baseline TS optimzers as seen in Figure \ref{fig:intend_count}. However, the single-inference model introduces a specific vulnerability: when the optimization trajectory enters regions of the PES that differ significantly from the training data, the predicted eigenvector may become inaccurate, leading to incorrect geometry updates. This leads to a higher rate of convergence failures compared to the baselines, as the single model lacks the ability to self-diagnose unreliable LMHE predictions (Figure \ref{fig:intend_count}).\par

To mitigate convergence failures caused by inaccurate eigenvector guidance, we implemented an ensemble consistency check which effectively neutralizes this failure mode. As illustrated in Supplementary Figure S1, predictions exhibiting higher ensemble variance $\sigma$ among five independently trained models are statistically more likely to be inaccurate. By utilizing this variance as a real-time proxy for uncertainty, the optimizer can dynamically detect unreliable predictions. When the ensemble disagreement $\sigma$ exceeds a predefined threshold, the optimizer triggers a fallback to an exact Hessian calculation with automatic differentiation at that step only. As observed in Supplementary Figure S2, the ensemble check is largely insensitive to the value of the variance threshold, and hence we utilize an intermediate value of 0.065.  As seen in Figure \ref{fig:intend_count}, the LMHE-ensemble approach now outperforms the baselines in reducing failure rates while retaining the high success count of TS optimizations using the single inference LMHE approach. 

\begin{figure}[H]
\centering
\includegraphics[width=0.495\textwidth]{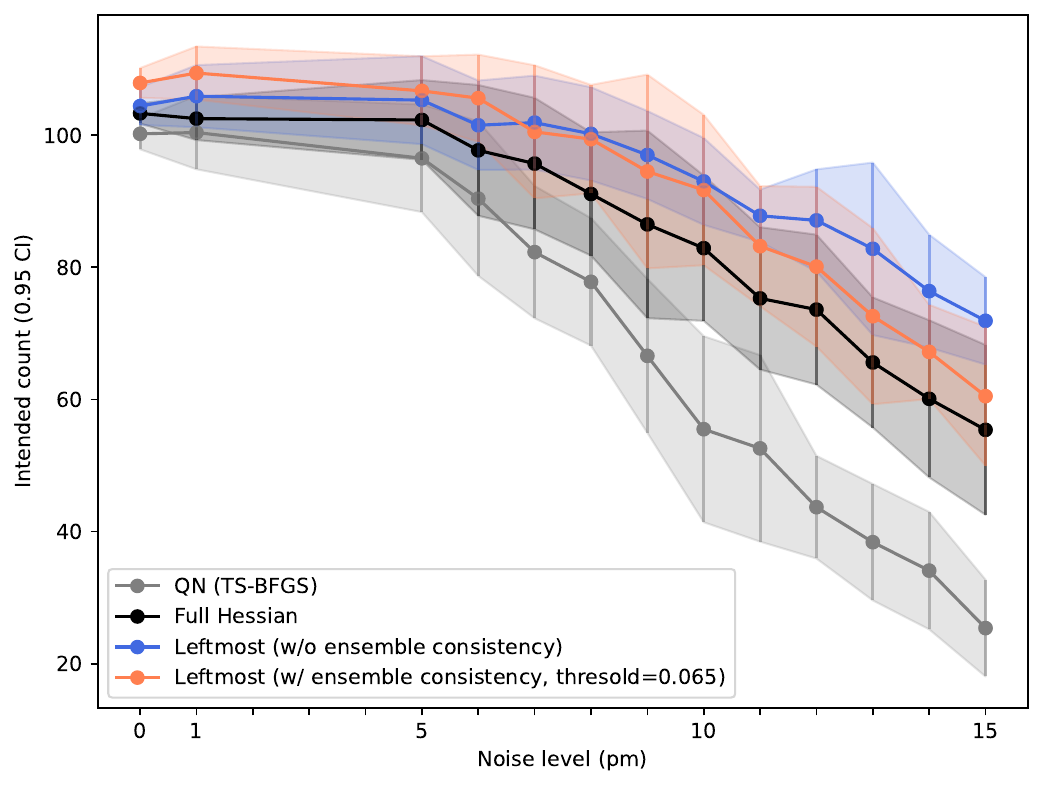}
\includegraphics[width=0.495\textwidth]{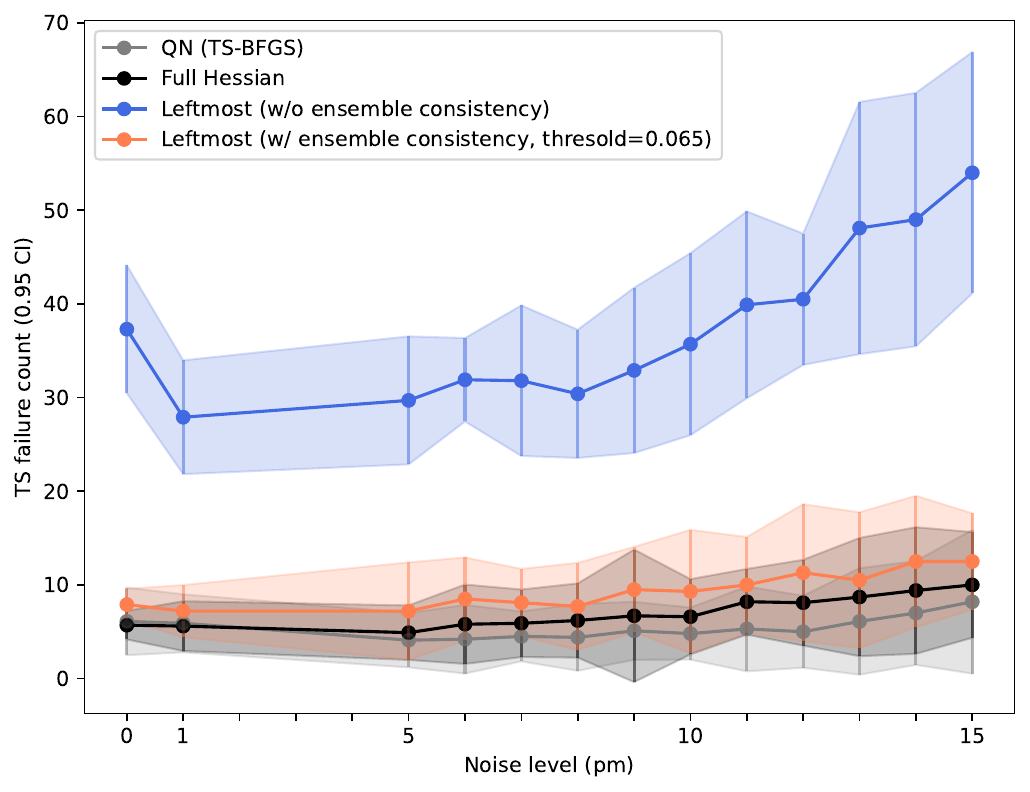}
 \caption{\textit{Comparison of the Robustness of TS Optimizers for 240 test combustion reactions.} (a) The count of intended TS optimizations is plotted against the amplitude of random Gaussian noise applied to the initial geometries. The LMHE optimizer exhibits  significantly more intended counts compared to the standard TS-BFGS QN method, maintaining high success rates comparable to the full Hessian approach even at high noise levels. (b) The number of TS optimization runs failing to converge is plotted against noise level. The single model LMHE strategy exhibits a higher failure rate due to inaccurate LMHE predictions. The LMHE ensemble consistency check identifies unreliable predictions, triggering a fallback to exact Hessian calculations and reducing failure rates to levels competitive with the full Hessian and QN baselines. Shaded regions represent the 95$\%$ confidence interval derived from 10 independent noise realizations.}
    \label{fig:noise_count}
\end{figure}

To investigate TS optimizer robustness, we systematically degraded the initial guess TS structures by introducing random Gaussian noise ranging from 0 to 15 pm to the Cartesian atomic positions to test the optimizers' capability to recover the intended TS from degraded starting geometries. Figure \ref{fig:noise_count}a shows that both the single inference and ensemble LMHE methods demonstrate robustness against structural noise comparable to the expensive full Hessian method for intended TS recovery rates. As the initial geometry degrades, the standard QN optimizer struggles to recover the correct ascent direction even with iterative diagonalization, leading to a rapid decline in the recovery rate for the intended TS. Even so Figure \ref{fig:noise_count}b shows that the single inference LMHE model still has higher failure rates, while the ensemble-LMHE approach is the best tradeoff for robustness and accuracy. As observed in Supplementary Figure S3, the ensemble check is largely insensitive to the value of the variance threshold, and hence again we utilize an intermediate value of 0.065. 

\begin{figure}[H]
\centering
\includegraphics[width=0.495\textwidth]{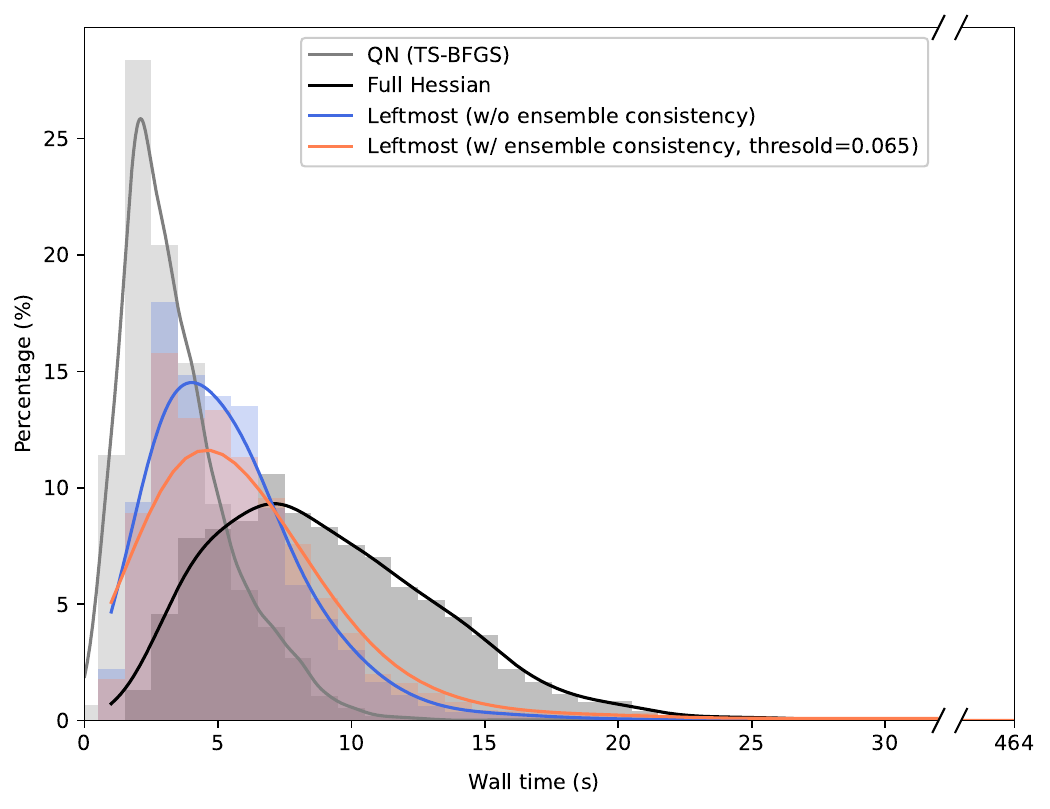}
\includegraphics[width=0.495\textwidth]{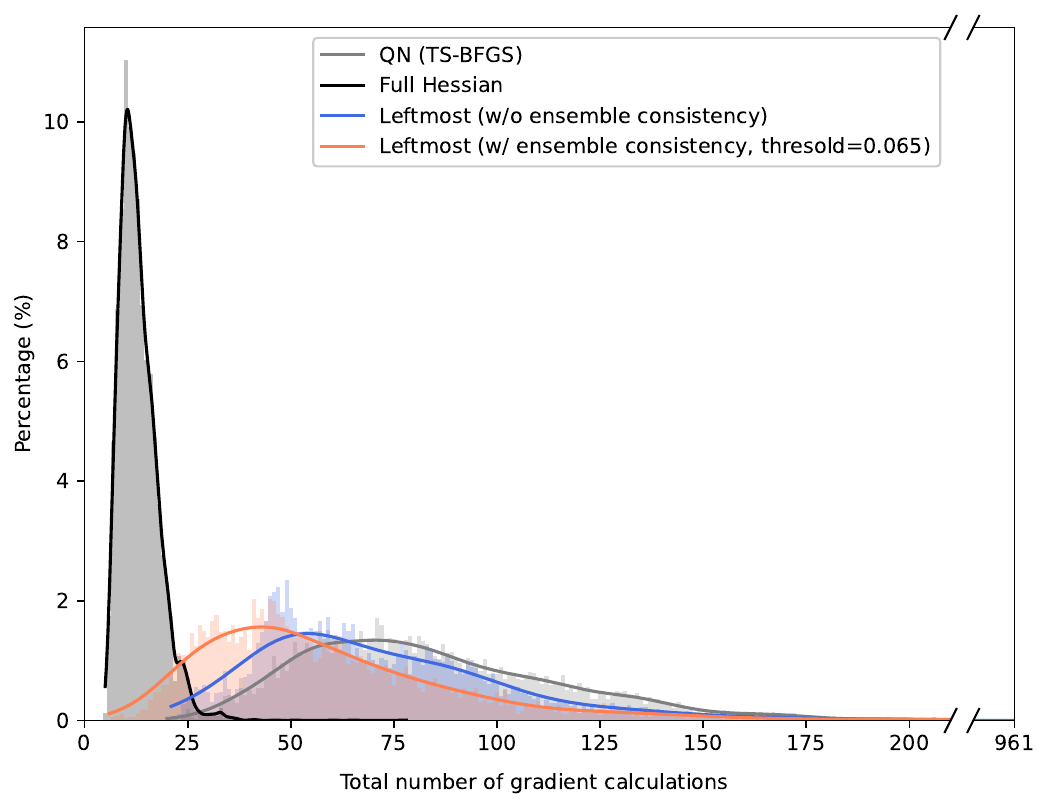}
    \caption{\textit{Wall time and number of gradient evaluation comparisons between TS optimizers.} (a) The distribution of wall times required for converged optimizations across all noise levels. While the standard QN method is the fastest, the full Hessian method suffers from a heavy tail due to the high cost of exact second-derivative calculations. The LMHE achieves a favorable balance, significantly reducing the computational cost relative to the full Hessian approach while providing the curvature information necessary for robust convergence compared to the QN baseline. (b) The distribution of total gradient calculations required for converged optimizations. While the full Hessian method requires the fewest evaluations due to exact curvature information, among the approximate methods both LMHE approaches require fewer total gradient evaluations than the standard QN method. To ensure comparability, the data includes only those optimization where all optimizers successfully recovered the intended TS from the same initial geometry.}
    \label{fig:wall_time}
\end{figure}

While the single inference and ensemble-LMHE matches the full Hessian's robustness, it does so at a fraction of the computational cost. Figure \ref{fig:wall_time}a illustrates the wall-time distribution for converged optimizations. The full Hessian method suffers from a "heavy tail," reflecting the expense of calculating second derivatives for every step. Both of our LMHE TS optimizers eliminates this tail, shifting the distribution significantly toward lower wall time values. The single-inference strategy defines the efficiency ceiling of our method, yielding the most compact time distribution by relying exclusively on rapid model predictions. Our ensemble-checked method effectively matches this performance, closely tracking the Single-Inference wall-time profile. This confirms that the ensemble-checked strategy does not simply recreate the full Hessian method via frequent fallbacks. Instead, it retains the high speed of the single-inference base for the vast majority of the trajectory, limiting expensive exact curvature calculations to rare, high-uncertainty regions. Finally, our method demonstrates a superior scalability regarding gradient evaluations, consistently reducing the total count of gradient evaluations compared to the QN baseline by eliminating the need for iterative diagonalization (Figure \ref{fig:wall_time}b). Although the standard QN method yields faster wall times, this occurs because the computational cost of the inference of our leftmost eigenvector predictor exceeds that of the inexpensive gradient evaluations on the MLIP PES used here. For completeness, Supplementary Figure S4, compares the computational cost for the ensemble check approach for various values of the variance threshold. 

\section{Conclusion}
We have presented a novel strategy for TS optimization that effectively bridges the gap between the robustness of second-order methods and the computational efficiency of first-order approaches. By directly predicting the LMHE which serves as the local approximation of the reaction coordinate, we demonstrated that a machine-learning model can provide the critical curvature information required for saddle point location without the prohibitive cost of DFT calculations nor the expense of automatic differentiation of the MLIP to create Hessians at each optimization step.\par

The central contribution of this study is the development of a robust optimizer that effectively integrates geometric deep learning with PRFO algorithms. Crucial to this success was addressing the geometric distinction between local atomic forces and the non-local nature of the LMHE. We showed that standard equivariant MPNNs, which rely on local diffusion of information, are insufficient for capturing the collective atomic motions characteristic of transition modes. The introduction of a scalable $E(3)$-equivariant global attention decoder was found to reduce both the prediction error and generalization gap by enabling the model to aggregate and broadcast information across the entire molecular graph.\par

Benchmarks on the Sella dataset reveal that LMHE optimization scheme is highly resilient to structural noise. Our approach recovers intended TS from degraded initial guesses with a success rate comparable to exact Hessian methods, yet does so with a wall-time distribution significantly faster than full Hessian methods. Furthermore, the implementation of an ensemble consistency check addresses the "black box" reliability issue common in machine learning. By using ensemble variance as a proxy for uncertainty, the optimizer dynamically identifies unreliable predictions and selectively falls back to the higher fidelity second order method, ensuring convergence stability. Ultimately, this methodology offers a scalable pathway for high-throughput reaction discovery, minimizing the need for human intervention in refining TS guesses.\par

\bibliography{references}

\bibliographystyle{unsrt}

\section*{Data Availability}
\noindent
All data \cite{data_avail} including the training, validation, and test datasets for the LMHE predictors, as well as the initial transition state guess structures, ground truth reactants and products, optimized transition states, and corresponding predicted reactants and products with their coordinates, are available at \url{https://doi.org/10.6084/m9.figshare.31791964}. Source data for Figs. \ref{fig:intend_count} to \ref{fig:wall_time} is available with this manuscript. Source data are provided in this paper. 

\section*{Code Availability}
\noindent
The codebase is comprised of several publicly available packages and tools that contribute to the project. The core framework developed for this study, including the implementation and training scripts for the GotenNet and GotenNet-GA models, model checkpoints, TS optimization scripts, and the Jupyter notebooks for evaluating ensemble consistency and analyzing optimization results, is accessible at \url{https://github.com/THGLab/LMHE-TSopt}. The TS optimizations were driven by the Sella optimization framework \cite{Sella2}, which is available with comprehensive documentation at \url{https://github.com/zadorlab/sella}. NewtonNet \cite{NewtonNet}, another integral part of the project, is also publicly available. Specifically, the HessianCalculator branch (\url{https://github.com/THGLab/NewtonNet/tree/HessianCalculator}) was utilized to ensure compatibility with the find-tuned models established in our previous work \cite{Yuan2024}. The parameters for this fine-tuned NewtonNet model \cite{Yuan2024} are publicly available at \url{https://github.com/THGLab/MLHessian-TSopt}. Finally,  our primary repository directly includes modified components from GotenNet \cite{GotenNet} with original code accessible at \url{https://github.com/sarpaykent/GotenNet}, as well as the spherical harmonics implementation extracted from the e3nn library \cite{e3nn} (\url{https://github.com/e3nn/e3nn}).

\section*{Acknowledgment}
\noindent
E.C.-Y.Y. and T.H-G. thank the CPIMS program, Office of Science, Office of Basic Energy Sciences, Chemical Sciences Division of the U.S. Department of Energy under Contract DE-AC02-05CH11231 for support. This work used computational resources provided by the National Energy Research Scientific Computing Center (NERSC), a U.S. Department of Energy Office of Science User Facility operated under Contract DE-AC02-05CH11231, and the Lawrencium computational cluster resource provided by the IT Division at the Lawrence Berkeley National Laboratory (Supported by the Director, Office of Science, Office of Basic Energy Sciences, of the U.S. Department of Energy under Contract No. DE-AC02-05CH11231).

\section*{Author contributions}
\noindent
G.W. and T.H.G. designed the LMHE method and wrote the manuscript. G.W. carried out all training and implemented and executed all workflows. All authors discussed the results and provided edits to the manuscript.

\section*{Competing interests}

The authors declare no competing interests.

\end{document}


\title{\textbf{Supplementary Information: Machine-Learned Leftmost Hessian Eigenvectors for Robust Transition State Finding}}

\author{Guanchen Wu$^{1}$, Chung-Yueh Yuan$^{1,5}$, Kareem Hegazy$^{4,7
}$, Samuel M. Blau$^{6}$, and Teresa Head-Gordon$^{1-3,5}$}
\date{}
\maketitle

\begin{center}
\vspace{-24pt}
$^1$Kenneth S. Pitzer Theory Center and Department of Chemistry, $^2$Department of Bioengineering, $^3$Department of Chemical and Biomolecular Engineering, $^4$Department of Statistics, University of California, Berkeley, CA, 94720 USA\\
$^5$Chemical Sciences Division and $^6$Energy Technologies Area, Lawrence Berkeley National Laboratory, Berkeley, CA, 94720 USA\\
$^7$International Computer Science Institute, Berkeley, CA 94704 USA\\

corresponding author: thg@berkeley.edu
\end{center}

\section*{Figures}
\begin{figure}[H]
\centering
\includegraphics[width=0.9\textwidth]{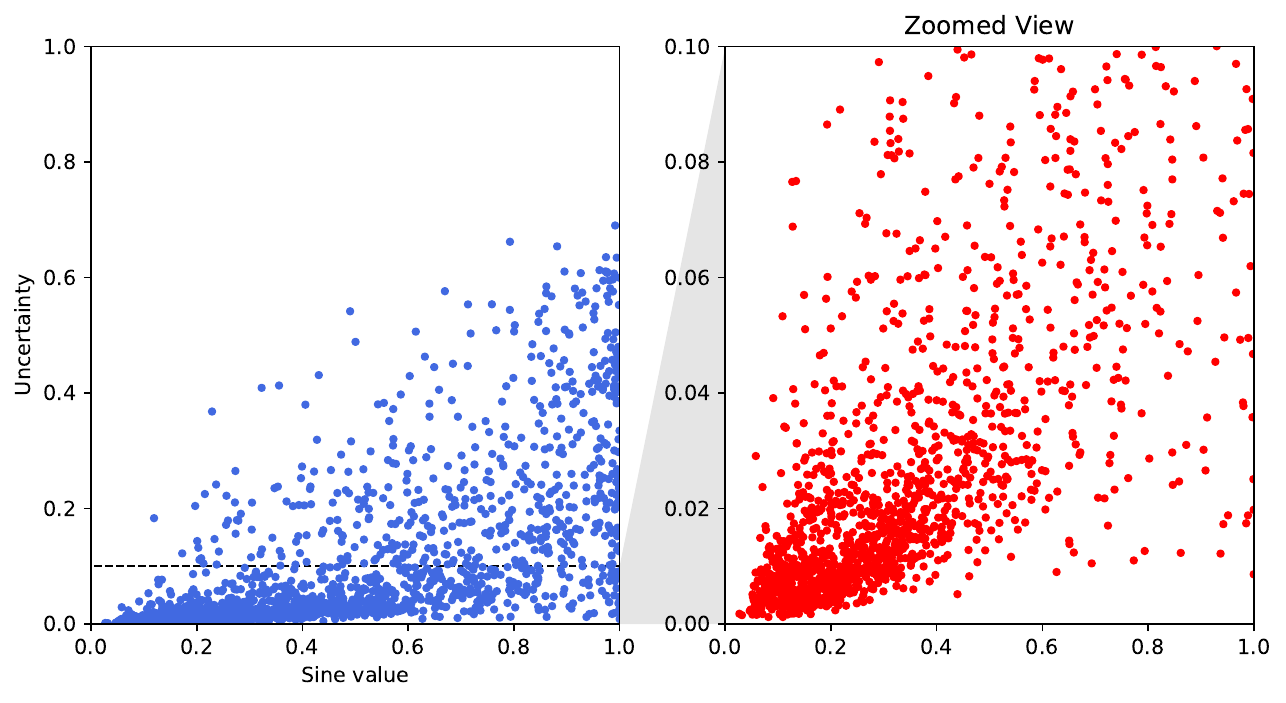}
    \caption{\textit{Correlation between uncertainty and prediction error.} A scatter plot of the ensemble uncertainty metric versus the sine value (error) of the predicted leftmost eigenvector relative to the ground truth. To maintain visual clarity, the results of 2,000 configurations were randomly sampled from the full augmented T1x test set. The positive correlation demonstrates that higher variance among the ensemble members serves as a reliable indicator of inaccurate predictions.}
    \label{fig:uncertainty}
\end{figure}

\begin{figure}[H]
\centering
\includegraphics[width=0.85\textwidth]{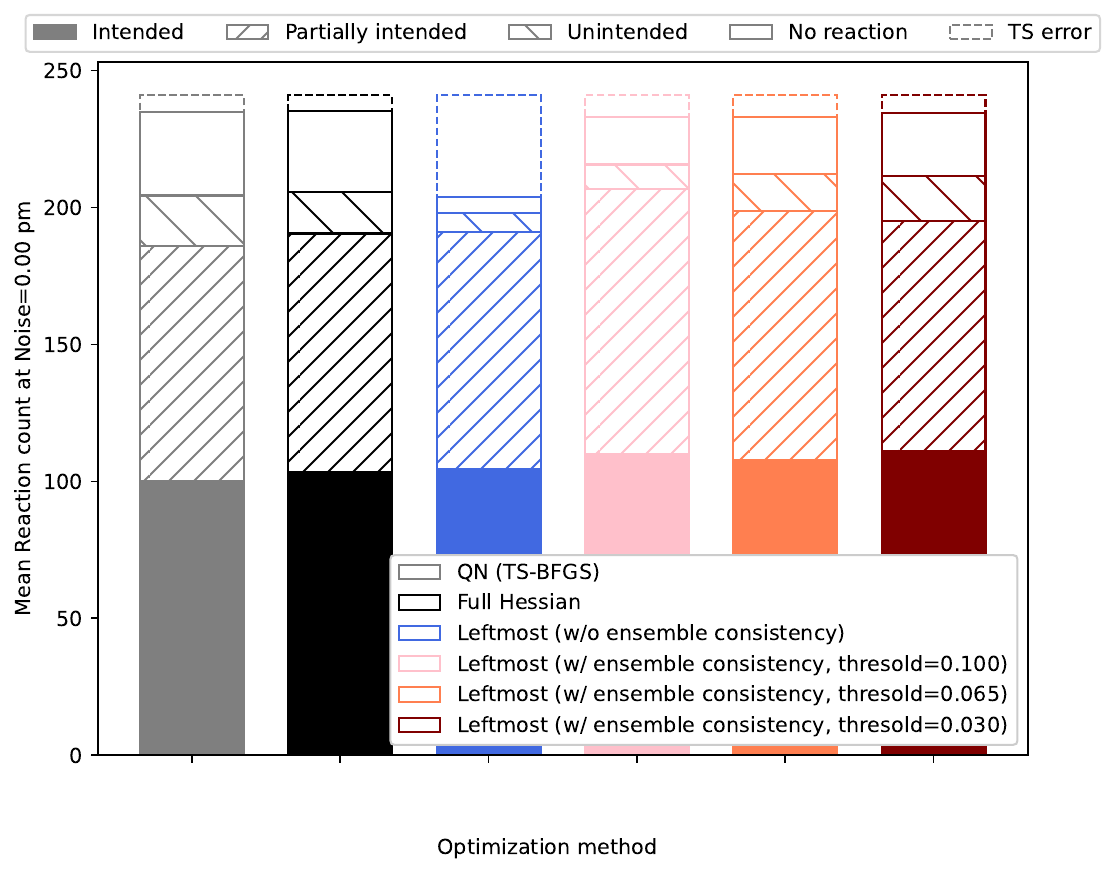}
  \caption{\textit{Ensemble check variation in variance threshold.} The single model LMHE strategy exhibits a higher failure rate due to inaccurate LMHE predictions. The LMHE ensemble consistency check identifies unreliable predictions, triggering a fallback to exact Hessian calculations and reducing failure rates to levels that are competitive with the performance of the full Hessian optimization, and significantly outperforming the standard QN baseline. The ensemble consistency approach is largely insensitive to values of the threshold. }
    \label{fig:intend_count}
\end{figure}

\begin{figure}[H]
\centering
\includegraphics[width=0.495\textwidth]{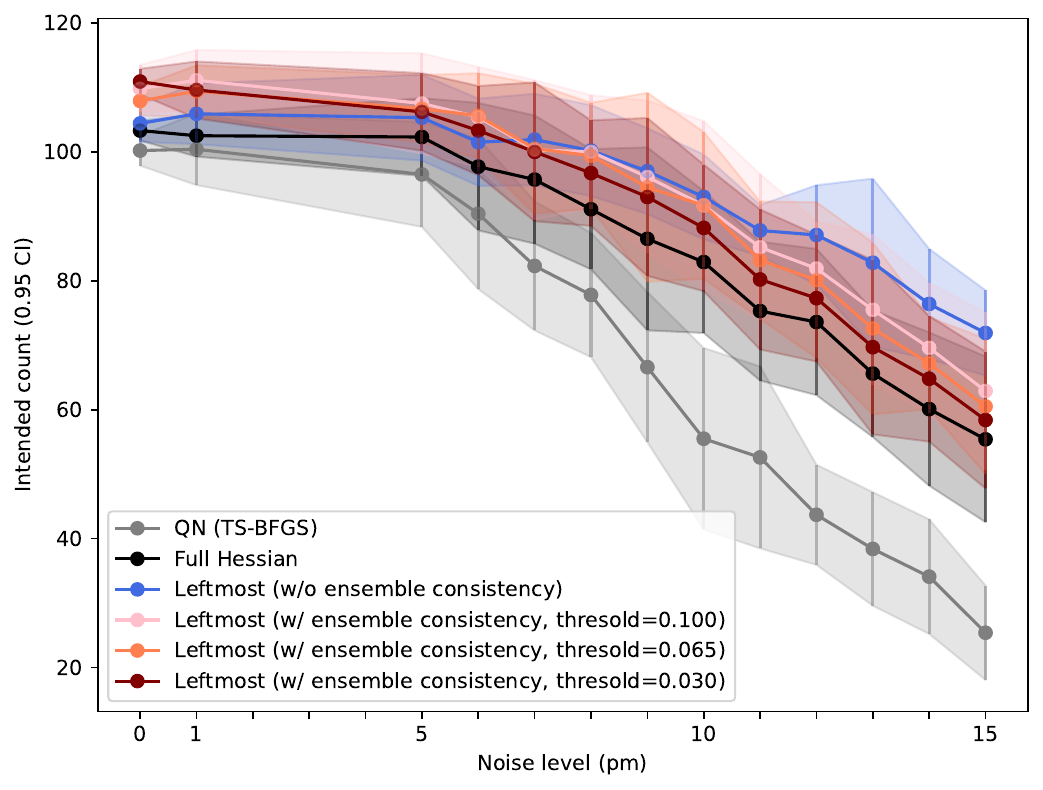}
\includegraphics[width=0.495\textwidth]{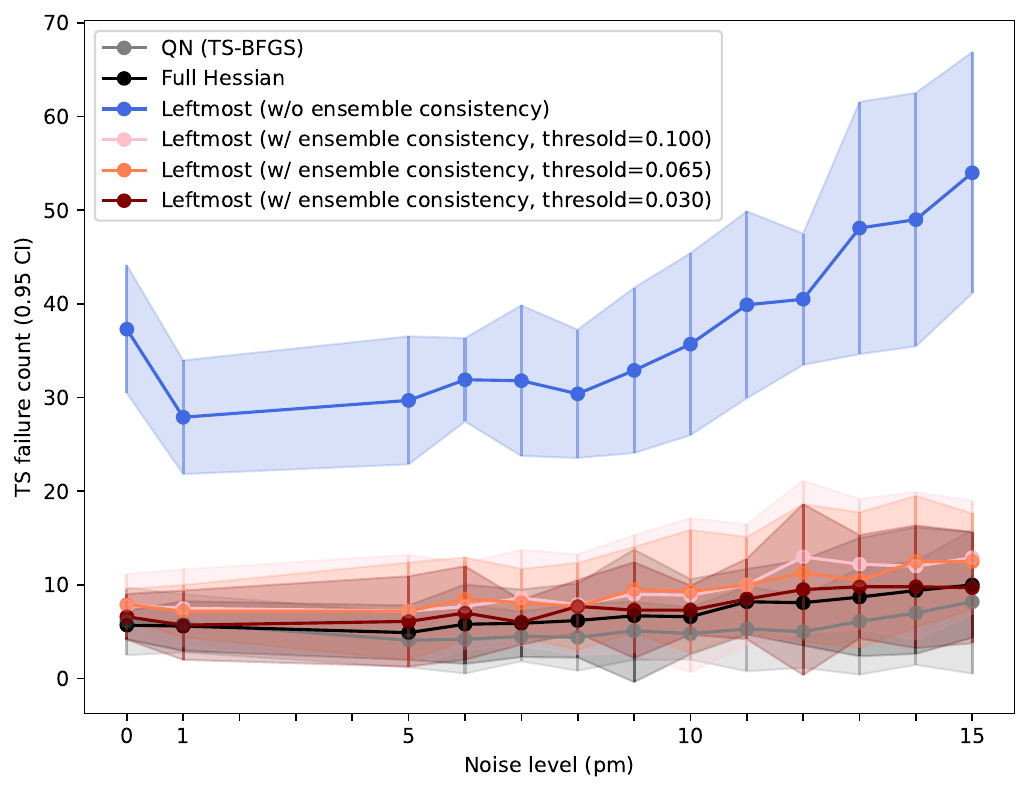}
 \caption{\textit{Comparison of the Robustness of TS Optimizers for 240 test combustion reactions.} (a) The count of intended TS optimizations is plotted against the amplitude of random Gaussian noise applied to the initial geometries. The LMHE optimizer exhibits  significantly more intended counts compared to the standard TS-BFGS QN method, maintaining high success rates comparable to the full Hessian approach even at high noise levels. (b) The number of TS optimization runs failing to converge is plotted against noise level. The single model LMHE strategy exhibits a higher failure rate due to inaccurate LMHE predictions. The LMHE ensemble consistency check identifies unreliable predictions, triggering a fallback to exact Hessian calculations and reducing failure rates to levels competitive with the full Hessian and QN baselines. Shaded regions represent the 95$\%$ confidence interval derived from 10 independent noise realizations.}
    \label{fig:noise_count}
\end{figure}

\begin{figure}[H]
\centering
\includegraphics[width=0.495\textwidth]{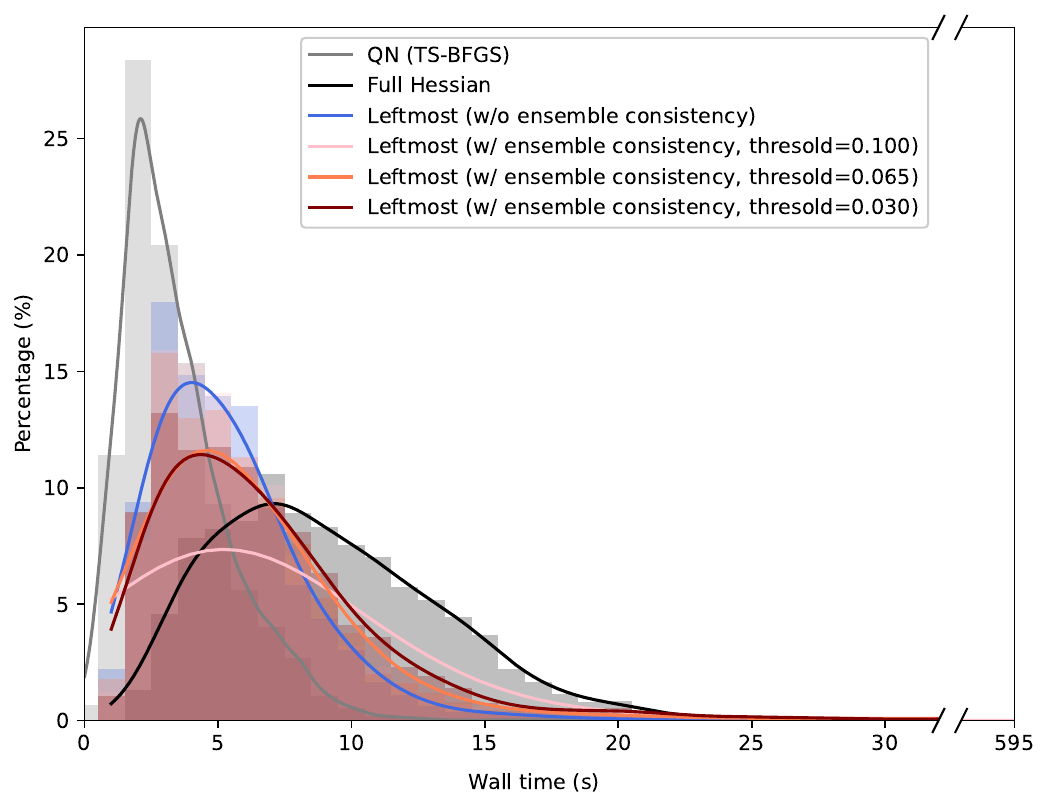}
\includegraphics[width=0.495\textwidth]{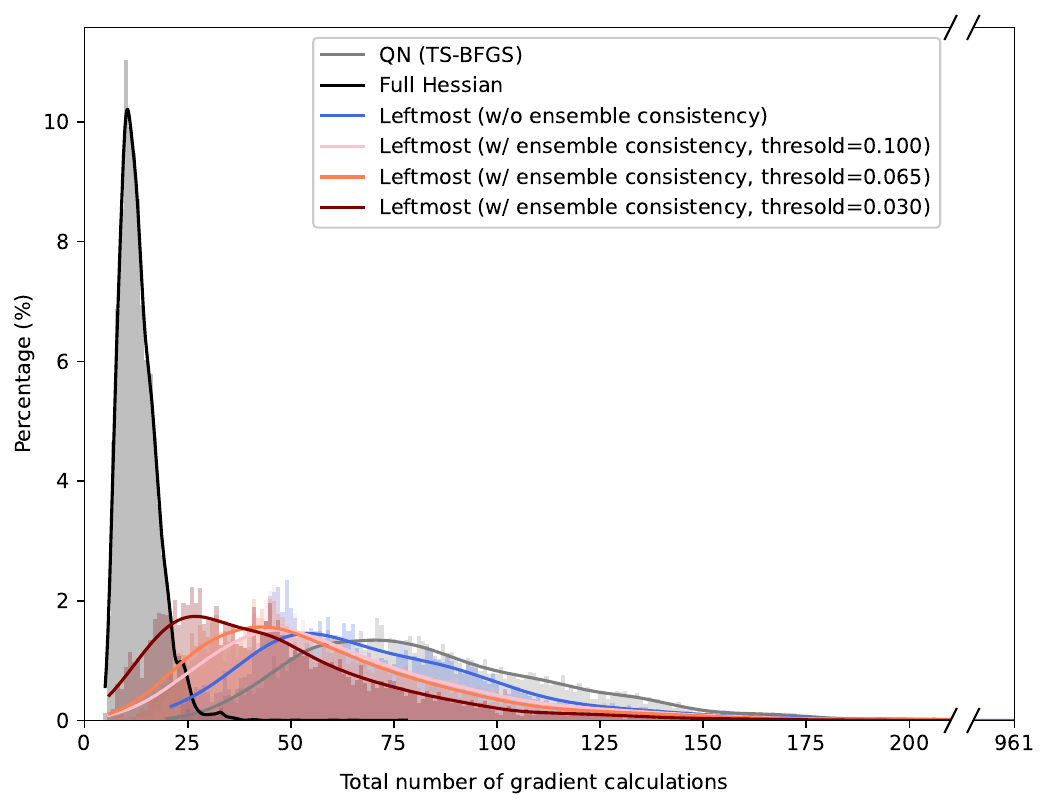}
    \caption{\textit{Wall time and number of gradient evaluation comparisons between TS optimizers.} (a) The distribution of wall times required for converged optimizations across all noise levels. While the standard QN method is the fastest, the full Hessian method suffers from a heavy tail due to the high cost of exact second-derivative calculations. The LMHE achieves a favorable balance, significantly reducing the computational cost relative to the full Hessian approach while providing the curvature information necessary for robust convergence compared to the QN baseline. (b) The distribution of total gradient calculations required for converged optimizations. While the full Hessian method requires the fewest evaluations due to exact curvature information, among the approximate methods both LMHE approaches require fewer total gradient evaluations than the standard QN method. To ensure comparability, the data includes only those optimization where all optimizers successfully recovered the intended TS from the same initial geometry.}
    \label{fig:wall_time}
\end{figure}

\section*{Tables}

\begin{table}[H] 
    \centering
    \caption{\textit{Hyper-parameters for the augmented T1x dataset.} The specific hyper-parameters used to train our encoder-decoder architecture (GotenNet-GA) and the baseline (GotenNet) models at two scales: Small (S, $\sim$2.1 M parameters) and Medium (M, $\sim$10.1 M parameters). The five independent models trained for the ensemble-based TS optimizer, including the model used for the single-model optimizer, all utilize the same hyper-parameters as the GotenNet-GA$_\textrm{M}$ model.} \label{t1x_hyperparameters}
    \begin{tabular}{lcccc}
        \toprule
         Hyper-parameters & GotenNet-GA$_{\text{S}}$ & GotenNet-GA$_{\text{M}}$ & GotenNet$_{\text{S}}$ & GotenNet$_{\text{M}}$ \\
        \midrule
        Optimizer & RAdamW & RAdamW & RAdamW & RAdamW \\
        Learning rate scheduling & \multicolumn{4}{c}{Reduce on plateau}   \\
        Maximum learning rate & 1e$-$4 & 1e$-$4 & 1e$-$4 & 1e$-$4  \\
        Learning rate decay & 0.8 & 0.8 & 0.8 & 0.7 \\
        Learning rate patience & 5 & 5 & 5 & 5 \\
        Loss function  & \multicolumn{4}{c}{$1-(\text{Cosine Similarity})^2$}  \\
        Gradient clipping & 5.0 & 5.0 & 5.0 & 5.0 \\
        Batch size  & 64 & 64 & 64 & 64 \\
        Early stopping  & \multicolumn{4}{c}{Minimizing validation loss}  \\
        Early stopping patience  & 25 & 25 & 25 & 25 \\
        Maximum number of epochs  & 1000 & 1000 & 1000 & 1000 \\
        Dropout rate  & 0.1 & 0.1 & 0.1 & 0.1 \\
        Node dimension ($d_{ne}$) & 128 & 256 & 128 & 256 \\
        GotenNet edge dimension & 128 & 256 & 128 & 256 \\
        GotenNet edge refinement dimension & 128 & 256 & 128 & 256 \\
        Maximum degree ($L_{max}$) & 2 & 2 & 2 & 2 \\
        Number of GotenNet layers & 2 & 4 & 4 & 5 \\
        Number of global attention layers & 2 & 1 & $-$ & $-$ \\
        Number of radial basis functions & 32 & 32 & 32 & 32 \\
        Number of inducing points ($M$) & 16 & 16 & $-$ & $-$ \\
        Number of Attention Heads & 8 & 8 & 8 & 8 \\
        Cutoff radius & 5.0 & 5.0 & 5.0 & 5.0 \\
         \bottomrule
     \end{tabular}
 \end{table}

\newpage

\section*{Appendices}

\noindent
\large
\textbf{Appendix 1. TS-BFGS Formulation} \label{qn_backbones}

\normalsize
\noindent
As introduced in Section 4.3 in the main text, the perpendicular component of the Hessian, $\mathbf{H}^{(k+1)}_{\perp}$, is updated using a standard QN formula. Generally, the QN update function takes the form $\mathbf{H}^{(k+1)} = \mathrm{QN}(\mathbf{H}^{(k)}, \mathbf{s}, \mathbf{y}).$, where $\mathbf{s}=\mathbf{x}^{(k+1)}-\mathbf{x}^{(k)}$ and $\mathbf{y}=\mathbf{g}^{(k+1)}-\mathbf{g}^{(k)}$. In our method, this update is applied in the projected subspace, i.e. $\mathbf{H}_{\perp}^{(k+1)} = \mathrm{QN}(\mathbf{H}_{\perp}^{(k)}, \mathbf{s}_{\perp}, \mathbf{y}_{\perp})$. For clarity in the descriptions of the standard algorithms in Appendix \ref{qn_backbones}, the $\perp$ subscript is omitted.\par

This work utilizes the TS-BFGS method \cite{tsbfgs1, tsbfgs2}. The formulation adopted in this work is given by:
\begin{align}
    |\mathbf{H}^{(k)}|&=\sum_i^d |\lambda_i^{(k)}| \mathbf{v}_i^{(k)} \mathbf{v}_i^{(k)^{\mathrm{T}}}, \\
    \mathbf{M}&=\mathbf{y}\mathbf{y}^{\mathrm{T}} + |\mathbf{H}^{(k)}| \mathbf{s} \mathbf{s}^{\mathrm{T}} |\mathbf{H}^{(k)}|, \\
    \mathbf{j}&=\mathbf{y}-\mathbf{H}^{(k)}\mathbf{s},\\
    \mathbf{u}&=\frac{\mathbf{M}\mathbf{s}}{\mathbf{s}^{\mathrm{T}} \mathbf{M}\mathbf{s}}, \\
    \mathbf{E}&= \mathbf{u}\mathbf{j}^{\mathrm{T}} + \mathbf{j}\mathbf{u}^{\mathrm{T}}-\mathbf{u}\mathbf{j}^{\mathrm{T}} \mathbf{s}\mathbf{u}^{\mathrm{T}},\\
    \mathbf{H}^{(k+1)}&=\mathbf{H}^{(k)}+ \mathbf{E},
\end{align}
where $d$ is the number of dimensions in $\mathbf{H}^{(k)}$, and $\lambda_i^{(k)}$ and $\mathbf{v}^{(k)}_i$ are the eigenvalues and eigenvectors of $\mathbf{H}^{(k)}$. \\

\noindent
\large
\textbf{Appendix 2. Global Attention Decoder Architecture} \label{decoder_architecture}

\normalsize
\noindent
As introduced in the main text, the decoder is composed of global attention layers designed to capture context in the entire molecule, built upon the induced set attention mechanism \cite{SetTransfromer}. As is done in GotenNet, we denote the tensors on each node into two types of features: scalar features $\mathbf{h}\in \mathbb{R}^{d_{ne}}$ invariant under transformations, and high-degree steerable features $\tilde{\mathbf{X}}^{(l)} \in \mathbb{R}^{d_{ne} \times (2l+1)}$ whose transformations depend on their degree $l$, where $d_{ne}$ denotes node embedding dimension.\par

In each induced set attention layer, there are $M$ inducing points, where $M$ is a hyper-parameter. Each inducing point consists of trainable scalar parameters $\mathbf{h}^{I} \in \mathbb{R}^{d_{ne}}$. First we aggregate information from graph nodes into inducing points through attention. For inducing point $\alpha$ and graph node $i$, we compute the query ($\mathbf{q}$) and key ($\mathbf{k}$) representations:
\begin{equation}
    \mathbf{q}_{\alpha}=\mathbf{W}_{qa}\mathbf{h}^{I}_{\alpha},\quad \mathbf{k}_i=\mathbf{W}_{ka}\mathbf{h}_{i},
\end{equation}
where $\mathbf{W}_{qa}, \mathbf{W}_{ka}\in \mathbb{R}^{d_{ne}\times d_{ne}}$ are learnable matrices. A softmax attention is subsequently applied:
\begin{equation}
    a_{\alpha i}=\frac{\exp(\mathbf{q}_{\alpha}^{\mathrm{T}} \mathbf{k}_i /\sqrt{d_{ne}} )}{\sum_{j\in \mathcal{N}} \exp(\mathbf{q}_{\alpha}^{\mathrm{T}} \mathbf{k}_j /\sqrt{d_{ne}})},
\end{equation}
where $\mathcal{N}$ is the set of all nodes in the graph. We compute the value ($\mathbf{v}$ and $\tilde{\mathbf{V}}^{(l)}$) representation:
\begin{equation}
    \mathbf{v}_{i}=\mathbf{W}_{va}\mathbf{h}_i, \quad \tilde{\mathbf{V}}_i^{(l)}=\mathbf{W}_{vva}^{(l)} \tilde{\mathbf{X}}_{i}^{(l)},
\end{equation}
where $\mathbf{W}_{va}, \mathbf{W}_{vva}^{(l)}\in \mathbb{R}^{d_{ne}\times d_{ne}}$ are learnable matrices. Here, $\mathbf{W}_{vva}^{(l)}$ is degree-specific, and uniform weights are applied across spatial dimensions under the same degree to preserve equivariance. The attention outputs are then computed:
\begin{equation}
    \mathbf{h}^{attn}_{\alpha} = \sum_{i\in \mathcal{N}} a_{\alpha i} \mathbf{v}_i, \quad
    \tilde{\mathbf{X}}_{\alpha}^{(l),\, attn}=\sum_{i\in \mathcal{N}} a_{\alpha i}\tilde{\mathbf{V}}_i^{(l)}.
\end{equation}
Subsequently, the features of the inducing points are updated:
\begin{equation}
    \mathbf{h}^{I}_{\alpha} \leftarrow \mathbf{h}^{I}_{\alpha}+\mathbf{W}_{oa} \mathbf{h}^{attn}_{\alpha},\quad \tilde{\mathbf{X}}^{(l),\,I}_\alpha \leftarrow \mathbf{W}_{voa} \tilde{\mathbf{X}}_{\alpha}^{(l),\, attn},
\end{equation}
where $\mathbf{W}_{oa}, \mathbf{W}_{voa} \in \mathbb{R}^{d_{ne}\times d_{ne}}$ are learnable matrices, and $\mathbf{W}_{voa}$ is shared across degrees. The updated features are then passed through an equivariant feed-forward (EQFF) layer \cite{GotenNet}:
\begin{equation}
    \mathbf{h}^{I}_{\alpha},\, \tilde{\mathbf{X}}^{(l),\,I}_\alpha \leftarrow \mathrm{EQFF}(\mathbf{h}^{I}_{\alpha},\, \tilde{\mathbf{X}}^{(l),\,I}_\alpha ).
\end{equation}\par

Second, we update the features on graph nodes with the aggregated information on inducing points through another attention layer. The query, key and value representations of graph node $i$ and inducing point $\alpha$ are computed:
\begin{equation}
\begin{split}
    \mathbf{q}_i=\mathbf{W}_{qb}\mathbf{h}_i, \quad \mathbf{k}_{\alpha}=&\mathbf{W}_{kb}\mathbf{h}_{\alpha}^I,\quad \mathbf{v}_{\alpha}=\mathbf{W}_{vb}\mathbf{h}_{\alpha}^I,\\
    \tilde{\mathbf{Q}}_i^{(l)}=\mathbf{W}_{vqb}\tilde{\mathbf{X}}_i^{(l)},\quad \tilde{\mathbf{K}}_\alpha^{(l)}=&\mathbf{W}_{vkb}^{(l)}\tilde{\mathbf{X}}^{(l),\,I}_\alpha, \quad \tilde{\mathbf{V}}_\alpha^{(l)}=\mathbf{W}_{vvb}^{(l)}\tilde{\mathbf{X}}^{(l),\,I}_\alpha,
\end{split}
\end{equation}
where $\mathbf{W}_{qb}, \mathbf{W}_{kb}, \mathbf{W}_{vb}, \mathbf{W}_{vqb}, \mathbf{W}_{vkb}^{(l)}, \mathbf{W}_{vvb}^{(l)}\in \mathbb{R}^{d_{ne}\times d_{ne}}$ are learnable matrices. $\mathbf{W}_{vqb}$ is shared across degrees while $\mathbf{W}_{vkb}^{(l)}$ and $\mathbf{W}_{vvb}^{(l)}$ are degree-specific. The softmax attention is subsequently computed:
\begin{equation} \label{attention_logit}
\begin{split}
    w_{i\alpha}&=\mathbf{q}_i^\mathrm{T}\mathbf{k_{\alpha}}+\sum_{l=1}^{L_{max}} {\tilde{\mathbf{Q}}_i^{(l)^\mathrm{T}}} \tilde{\mathbf{K}}_\alpha^{(l)},\\
    a_{i\alpha}&=\frac{\exp(w_{i\alpha} /\sqrt{(L_{max}+1)\cdot d_{ne}} )}{\sum_{\beta\in \mathcal{P}} \exp(w_{i\beta} /\sqrt{(L_{max}+1)\cdot d_{ne}})},
\end{split}
\end{equation}
where $\tilde{\mathbf{Q}}_i^{(l)}$ and $\tilde{\mathbf{K}}_\alpha^{(l)}$ are flattened into one dimensional vectors, $L_{max}$ is a hyper-parameter defining the maximum degree we include in the model, and $\mathcal{P}$ is the set of all inducing points. The attention output is then computed:
\begin{equation}
    \mathbf{h}^{attn}_{i} = \sum_{\alpha\in \mathcal{P}} a_{i\alpha} \mathbf{v}_\alpha, \quad
    \tilde{\mathbf{X}}_{i}^{(l),\, attn}=\sum_{\alpha\in \mathcal{P}} a_{i\alpha}\tilde{\mathbf{V}}_\alpha^{(l)}.
\end{equation}
Subsequently, the features of the graph nodes:
\begin{equation}
    \mathbf{h}_{i} \leftarrow \mathbf{h}_{i}+\mathbf{W}_{ob} \mathbf{h}^{attn}_{i},\quad
    \tilde{\mathbf{X}}^{(l)}_i \leftarrow \tilde{\mathbf{X}}_i^{(l)} + \mathbf{W}_{vob} \tilde{\mathbf{X}}_{i}^{(l),\, attn},
\end{equation}
where $\mathbf{W}_{ob}, \mathbf{W}_{vob} \in \mathbb{R}^{d_{ne}\times d_{ne}}$ are learnable matrices, and $\mathbf{W}_{vob}$ is shared across degrees. The updated features are then passed through an equivariant feed-forward (EQFF) layer \cite{GotenNet}:
\begin{equation}
    \mathbf{h}_{i},\, \tilde{\mathbf{X}}^{(l)}_i \leftarrow \mathrm{EQFF}(\mathbf{h}_{i},\, \tilde{\mathbf{X}}^{(l)}_i).
\end{equation}

\noindent
\large
\textbf{Appendix 3. Equivariance Proof of GotenNet-GA Model} \label{equivariance_proof}

\normalsize
\noindent
We demonstrate that the proposed architecture is equivariant under the Euclidean group $E(3)$. The goal is to show that the model's output, the LMHE $\mathbf{v}_1$ transforms physically under the group actions, being invariant to translation and equivariant to orthogonal transformations (rotations and reflections). Since the GotenNet encoder is inherently $E(3)$-equivariant, we focus this proof on demonstrating that the global attention decoder preserves these symmetry properties. The global attention decoder is constructed exclusively from four operations: linear projections, attention mechanisms, feature aggregation, and EQFF blocks. The EQFF block is adopted from GotenNet and its equivariance has been rigorously established in prior work \cite{GotenNet}. Therefore, it suffices to demonstrate the equivariance of the remaining three operations: linear projections, attention, and aggregation.\par

\textbf{Translation Invariance}. The GotenNet encoder generates node features $\mathbf{h}_i$ and $\tilde{\mathbf{X}}_i^{(l)}$ derived exclusively from relative coordinate vectors $\mathbf{r}_{ij}=\mathbf{r}_i-\mathbf{r}_j$ and atomic numbers \cite{GotenNet}, both remaining invariant under a global translation. Consequently, the input features to the decoder are invariant with respect to global translation. Since the decoder operations operate solely on these internal features without re-introducing absolute coordinates, the outputs of the decoder inherits this translation invariance.\par

\textbf{Orthogonal Equivariance}. We next prove equivariance under the orthogonal group $O(3)$. Let $g\in O(3)$ be an arbitrary orthogonal transformation. The scalar features $\mathbf{h}$ can be considered as zero-degree features $\tilde{\mathbf{X}}^{(0)}$. For the features $\tilde{\mathbf{X}}^{(l)}$, we employ a flattened column vector notation $\mathbf{x}^{(l)}\in \mathbb{R}^{d_{ne}(2l+1)}$ to facilitate the analysis. The group action on this flattened vector is defined by the block-diagonal matrix $\mathcal{D}^{(l)}(g)$ :
\begin{equation}
    \mathcal{D}^{(l)}(g)=\mathbf{I}_{d_{ne}}\otimes \mathbf{D}^{(l)}(g),
\end{equation}
where $\mathbf{I}_{d_{ne}}$ is the identity matrix of dimension $d_{ne}$ and $\mathbf{D}^{(l)}(g)$ is the Wigner D-matrix of degree $l$ for the transformation $g$ (accounting for parity if $g$ includes reflection). The transformed feature vector is $\mathbf{x}^{(l)'} = \mathcal{D}^{(l)}(g)\mathbf{x}^{(l)}$.\par

\textbf{Linear Projections:} The decoder employs degree-specific linear projections parameterized by weights $\mathbf{W}\in \mathbb{R}^{d_{ne}\times d_{ne}}$. In our vectorized formalism, this operation is expressed as $\mathcal{W}=\mathbf{W}\otimes \mathbf{I}_{2l+1}$. We verify that this operation commutes with the group action:
\begin{equation}
\begin{split}
      \mathcal{D}^{(l)}(g)\mathcal{W}
    &=(\mathbf{I}_{d_{ne}}\otimes \mathbf{D}^{(l)}(g))(\mathbf{W}\otimes \mathbf{I}_{2l+1})\\
    &=(\mathbf{I}_{d_{ne}}\mathbf{W})\otimes(\mathbf{D}^{(l)}(g)\mathbf{I}_{2l+1})\\
    &=(\mathbf{W}\mathbf{I}_{d_{ne}})\otimes(\mathbf{I}_{2l+1}\mathbf{D}^{(l)}(g))\\
    &=(\mathbf{W}\otimes \mathbf{I}_{2l+1})(\mathbf{I}_{d_{ne}}\otimes \mathbf{D}^{(l)}(g))\\
    &=\mathcal{W}\mathcal{D}^{(l)}(g).
\end{split}
\end{equation}
Since the operators commute, the projected query ($\mathbf{q}$), key ($\mathbf{k}$), and value ($\mathbf{v}$) vectors transform equivariantly according to $\mathcal{D}^{(l)}(g)$.\par

\textbf{Attention Mechanism}: The attention logits (Eq. \ref{attention_logit}) utilize the inner product of flattened feature vectors of each degree $\mathbf{q}^{(l)^\mathrm{T}}\mathbf{k}^{(l)}$. Under the transformation $g$:
\begin{equation}
\begin{split}
      (\mathbf{q}^{(l)'})^\mathrm{T}\mathbf{k}^{(l)'}
    &=(\mathcal{D}^{(l)}(g)\mathbf{q}^{(l)})^{\mathrm{T}}(\mathcal{D}^{(l)}(g)\mathbf{k}^{(l)})\\
    &=\mathbf{q}^{(l)^{\mathrm{T}}}\mathcal{D}^{(l)}(g)^{\mathrm{T}}\mathcal{D}^{(l)}(g)\mathbf{k}^{(l)},
\end{split}
\end{equation}
where matrix $\mathcal{D}^{(l)}(g)$ is orthogonal:
\begin{equation}
    \mathcal{D}^{(l)}(g)^{\mathrm{T}}\mathcal{D}^{(l)}(g)=\mathbf{I}_{d_{ne}} \otimes (\mathbf{D}^{(l)}(g)^{\mathrm{T}}\mathbf{D}^{(l)}(g))=\mathbf{I}_{d_{ne}} \otimes \mathbf{I}_{2l+1}=\mathbf{I}.
\end{equation}
Thus, the inner product is preserved, ensuring that the attention weights $a$ remain invariant scalars.\par

\textbf{Feature Aggregation:} The attention output is a linear combination of value vectors weighted by invariant attention weights. The aggregated output under transformation $g$ is:
\begin{equation}
    \mathbf{x}^{(l),attn'}=\sum a(\mathcal{D}^{(l)}(g)\mathbf{v}^{(l)})=\mathcal{D}^{(l)}(g)\left(\sum a \mathbf{v}^{(l)} \right)=\mathcal{D}^{(l)}(g)\mathbf{x}^{(l),attn}.
\end{equation}
This confirms that the output features of the global attention preserve $O(3)$-equivariance.

\bibliography{references}

\bibliographystyle{unsrt}